\documentclass[prx, amsmath, amssymb, superscriptaddress, twocolumn]{revtex4-1}

\usepackage[utf8]{inputenc}
\usepackage{graphicx}   
\usepackage{dcolumn}  
\usepackage{bm}          
\usepackage{xcolor}    
\usepackage{lineno} 
\usepackage{lipsum} 

\begin{document}

\title{The role of crosslinking density in surface stress and surface energy of soft solids}
\author{Weiwei Zhao}\affiliation{Department of Physics, The Hong Kong University of Science and Technology, Hong Kong, China}
\author{Jianhui Zhou}\affiliation{Department of Physics, The Hong Kong University of Science and Technology, Hong Kong, China}
\author{Haitao Hu}\affiliation{Department of Physics, The Hong Kong University of Science and Technology, Hong Kong, China}
\author{Chang Xu}\affiliation{School of Physical Science, University of Science and Technology of China, Hefei, China}

\author{Qin Xu}
\email{qinxu@ust.hk}
\affiliation{Department of Physics, The Hong Kong University of Science and Technology, Hong Kong, China}
\affiliation{HKUST Shenzhen Research Institute, Shenzhen, China}

\begin{abstract}
Surface stress and surface energy are two fundamental parameters that determine the surface properties of any materials. While it is commonly believed that the surface stress and surface energy of  liquids are identical, the relationship between the two parameters in soft polymeric gels remains debatable. In this work, we measured the surface stress and surface energy of soft silicone gels with varying weight ratios of crosslinkers in soft wetting experiments. Above a critical density, $k_0$, the surface stress was found to increase significantly with crosslinking density while the surface energy remained unchanged. In this regime, we can estimate a non-zero surface elastic modulus that also increases with the ratio of crosslinkers. By comparing the surface mechanics of the soft gels with their bulk rheology, the surface properties near the critical density $k_0$ were found to be closely related to the underlying percolation transition of the polymer networks.
\end{abstract}
\date{\today}
\maketitle

\section{Introduction}
Surface stress and surface energy are the essential parameters in many mechanical problems involving material interfaces, including adhesion and wetting between materials  \cite{Andreotti2020, Dufresne_annual_review2017,Chung2005}, the fracture formations dynamics \cite{Long2021,Creton_2016,Liu2020}, and the evolution of phase separations in composite systems \cite{Style2018,Xufeng2020}. While surface stress, $\Upsilon$, indicates the force per unit length required to expand a region at a material interface, surface energy, $\gamma$, characterizes the associated energy cost per unit area. At a liquid interface, molecules can easily redistribute themselves under deformation to keep a constant density such that the surface stress and surface energy are always identical, $\Upsilon = \gamma$.  Conventionally, the surface stress of a liquid is often referred to as liquid surface tension. By contrast, the surface densities of molecules or atoms in crystalline solids will vary through deformations. As a result, the surface stress and surface energy of crystalline solids can differ greatly, $\Upsilon\neq\gamma$ \cite{orowan1970, Cammarata1994,Streitz1994}.

However, despite the growing interest in the mechanics of soft polymeric gels, there is little consensus on whether their surface stress ($\Upsilon_g$) and surface energy ($\gamma_g$) are equal \cite{Liang2018,Xu2018,Snoeijer2018,Shih2019,Liang2018_Langumir}. In experiments with liquid droplets wetting on soft gels, for example, the macroscopic contact angle was found to remain constant as the substrate stretches up to $100~\%$ \cite{Schulman2018}. This finding implies a similarity between the surfaces of liquids and gels, such that the surface energy is strain-independent and hence consistently equal to the surface stress.  On the other hand, the direct imaging of local wetting profiles on the scales of tens of micrometers showed that the surface stress of soft gels can differ substantially from the surface energy \cite{Style2013,Xu2017}. Under highly asymmetric strain fields, the surface stress of soft gels can even be anisotropic, like crystalline solids \cite{Xu2018,Katrina2021}. However, despite the apparent discrepancies in the surface properties of soft solids among different studies, quantitative studies on the relationship between the surface stress and surface energy of soft polymeric gels are still lacking.

To address this issue, we systematically studied how the wetting of liquid droplets on soft gels is affected by the crosslinking density of the substrates. By measuring the droplet shapes and the substrate profiles separately on different length scales, we observed a gradual crossover from a solid-like regime where surface stress is greater than surface energy ($\Upsilon_g > \gamma_g$) to a liquid-like state where the two parameters become approximately equal ($\Upsilon_g\approx \gamma_g$) near a critical density of crosslinkers. We show that this transition in surface properties is physically related to an underlying change in the material rheology.  

\section{Experimental Results}
The gel substrates used in this work were prepared by mixing the divinyl-terminated polydimethylsiloxane (Gelest, DMS-V31) with a trimethylsiloxane terminated-dimethylsiloxane copolymer as the cross-linkers (Gelest, HMS-301) and a platinum-divinyltetramethyldisiloxane complex in xylene as the catalyst (Gelest, SIP6831.2) \cite{Jensen2015}. The pre-cured solution was spin-coated on standard $1.5$ thickness cover-slips at a speed of 800 rpm for one minute, and then cured at room temperature for about 40 hours before measurements. This preparation protocol yielded a gel layer approximately 56 $\mu m$ thick with a surface roughness around $20$ nm \cite{Xu2021}. The weight ratio of the crosslinkers, $k$, determines the stiffness of the substrates. In this work, we kept $k$ in the range of $0.7\%\sim1.4\%$ so that the resulting shear modulus could be adjusted between the orders of $10^1 \sim10^3$ Pa.

\begin{figure}
    \centering
    \includegraphics[width=85mm]{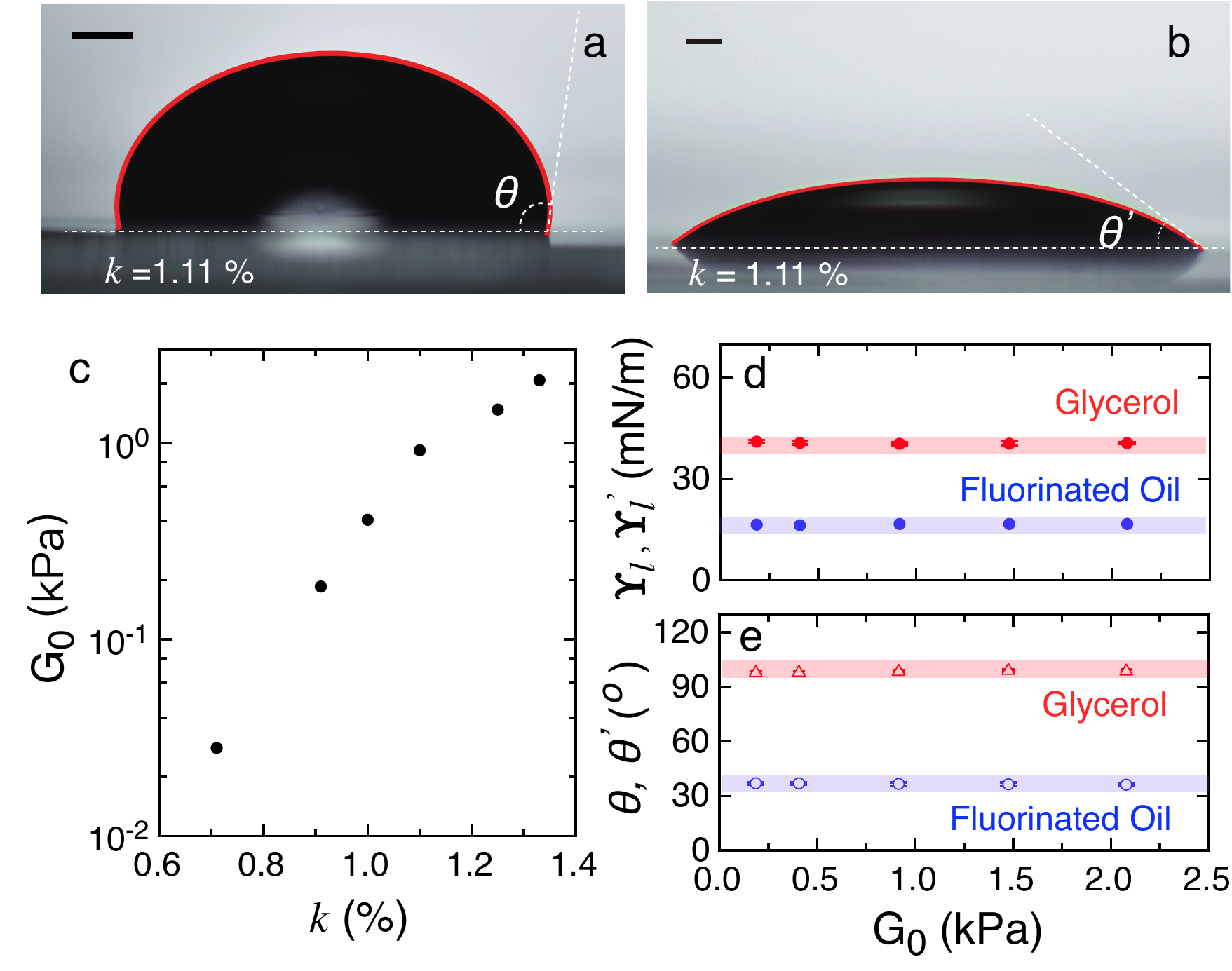}
    \caption{ {\bf Measurements of macroscopic contact angles.} \textbf{(a-b)} Snapshots of a glycerol droplet (a) and a fluorinated oil droplet (b) wetting the substrates of soft silicone gels with the crosslinking density $k= 1.11\%$. The dashed lines in the images indicate the macroscopic contact angle $\theta$. The solid red lines are the best fits to the drop boundaries with the measured surface stresses at the liquid-air interfaces. The scale bars  are 500 $\mu$m long in both the panels. \textbf{(c)} A plot of the shear modulus $G_0$ against the crosslinking density, $k$. \textbf{(d)} The surface stresses of the glycerol ($\Upsilon_l$) and fluorinated oil ($\Upsilon_l^\prime$) droplets  extracted from the wetting images (such as those in (a) and (b)) are plotted against the shear modulus $G_0$ of the substrates. The red-grey and blue-grey regions indicate the average values, $\Upsilon_l = 40.7\pm0.7$ mN/m and $\Upsilon_l^\prime = 16.7 \pm 0.6$ mN/m, respectively. \textbf{(e)} The plots of the contact angle against the shear modulus of the substrates $G_0$ for glycerol ($\theta$) and fluorinated oil ($\theta^\prime$) droplets, respectively. The red-grey and blue-grey regions represent the average values of measured contact angles, $\theta= 98.4^\circ\pm1.2^\circ$ and $\theta^\prime = 36.7^\circ\pm1.1^\circ$, for the two different deposited liquids.}
    \label{macroscopic}
\end{figure}

We first quantified how the macroscopic wetting profiles of liquid droplets were affected by the crosslinking densities of the gel substrates. We deposited millimeter-sized liquid droplets on the substrates using a pipette and then waited for 40 mins to ensure the wetting was in equilibrium. For all of the droplets measured in the experiments, we did not observe any wetting hysteresis on the soft gels \cite{Xu2017, Xu2018}. Figures \ref{macroscopic}a shows a representative image of a glycerol droplet wetting soft gels with the weight ratio of crosslinkers $k =1.11\%$. The deposited glycerol droplets had typical sizes of $3\sim 4$ mm, larger than the liquid capillary length ($\sim 1$ mm). Therefore,  the overall shapes of the droplets resulted from the balance between the gravitational stress and the Laplace pressure induced at liquid-air interfaces. For a liquid droplet with a given surface stress $\Upsilon_l$ and a mass density $\rho_l$, the stress balance can be expressed as: 
\begin{equation} 
2\Upsilon_l (\kappa-\kappa_0)=\rho_l g z
\label{balance}
\end{equation}
where $\kappa$ is the mean curvature of the liquid-air interface and $\kappa_0$ specifically represents the mean curvature at the apex. By solving the stress equation numerically and fitting the outcomes to the droplet boundaries, we can determine the surface stress of deposited droplets on each gel substrate \cite{HANSEN1991}. 

We measured the surface stresses of glycerol droplets as the crosslinking densities of the soft substrates ($k$) varied systematically. The shear modulus of the substrates $G_0$ increased sharply from 28 Pa to 2.1 kPa with increasing $k$ from $0.71~\%$ to $1.33~\%$ (Fig. \ref{macroscopic}c). For $k<0.7~\%$, the pre-mixed solution will have not cured properly but will have remained mainly as a fluid. As shown in Fig.~\ref{macroscopic}d, the surface stress of the deposited glycerol droplet remained constantly around $\Upsilon_{l} =40.7\pm 0.7$ mN/m while $G_0$ varied from 180 Pa to 2.1 kPa. It is noticeable that this constant value was considerably lower than the surface stress of pure liquid glycerol in air ($\sim 67$ mN/m), a phenomenon that has been observed in previous experiments \cite{Xu2017,Style2013}. This reduction in the surface stress was caused by uncrosslinked polymer chains that were extracted from the gel substrates. The extracted polymers covered the droplet surfaces and lowered the resulting surface stress \cite{Xu2021,Sebastien2018,Zhao2018,aurelie2017,Wong2020}. The modulus-independent surface stress $\Upsilon_{l}$ indicates that the extracted free chains were fully saturated at the liquid-air interfaces.

We next focus on the contact angles of the glycerol droplets $\theta$ on the same gels. For droplets much larger than the elastocapillary length of the substrates ($l_e\sim10^1$ $\mu$m), the equilibrium  $\theta$ follows the classical Young Dupre's law \cite{Style_PNAS_2013},   
\begin{equation}
    \cos{\theta}=\frac{\gamma_{ga}-\gamma_{gl}}{\gamma_l},
    \label{Young}
\end{equation}
where $\gamma_{ga}$ and $\gamma_{gl}$ are the surface energies of the soft gels interfacing with air and liquid, respectively. For liquid glycerol droplets, the surface energy is expected to be equal to the surface stress, $\gamma_{l} = \Upsilon_{l} =40.7$ mN/m. By resolving the droplet boundaries using imaging analysis, contact angles can be measured precisely. The red triangles in Fig.~\ref{macroscopic}g show that the contact angle of glycerol droplets was consistently around $\theta= 98.4^\circ \pm 1.2^\circ$ as the substrate modulus varied from $G_0=0.18$ kPa to $G_0=2.1$ kPa. According to Eq.~\ref{Young}, this result suggests that $\gamma_{ga}-\gamma_{gl}$ is independent of $G_0$ for glycerol droplets wetting on soft silicone gels. 

To test the generality of these findings, we further replaced glycerol with fluorinated oil (Sigma Aldrich FC-70) as a different deposited liquid. As shown in Fig. \ref{macroscopic}b, the fluorinated oil wetted the gel substrates well with a contact angle significantly smaller than 90 degrees. By varying the crosslinking density from $k=0.91\%$ to $k=1.33\%$, the surface stress of fluorinated oil droplets, $\Upsilon_{l}^\prime = \gamma_{l}^\prime =16.7\pm 0.6$ mN/m (Fig. \ref{macroscopic}d), and their wetting contact angles on the gels, $\theta^\prime = 36.7^\circ \pm 1.1^\circ$ (Fig. \ref{macroscopic}e), were both found to be independent of the substrate modulus as well. Since it is very unlikely that both $\gamma_{ga}$ and $\gamma_{gl}$ in Eq. \ref{Young} varied with $G_0$ in the exact same way for the two different deposited liquids (glycerol and fluorinated oil),  we conclude that the crosslinking density had no effect on the surface energy of the soft silicone gels.

While the equilibrium contact angle results from minimization of the overall surface energy, it provides little information on the deformations of the gel surfaces induced by wetting. It has been shown that the local profiles near the contact points are related to the gel surface stress \cite{Style2013}. Since surface stress ($\Upsilon_g$) is not necessarily equal to surface energy ($\gamma_g$) for polymeric gels, we need to quantify the wetting profiles microscopically to evaluate the influence of crosslinking density on the surface stress.

\begin{figure}[t]
    \centering
    \includegraphics[width=85mm]{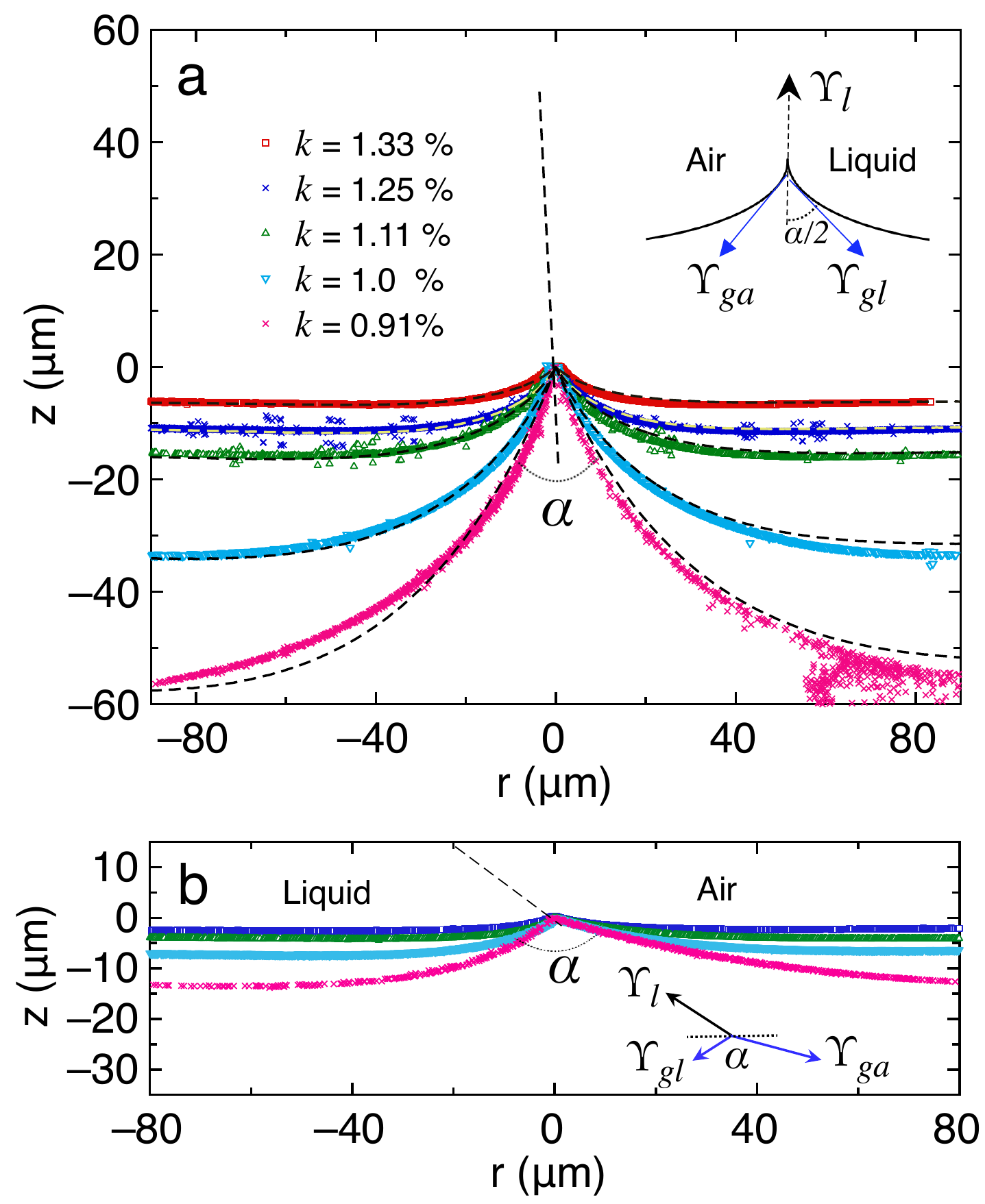}
    \caption{{\bf Microscopic wetting profiles on differently crosslinked gel substrates.} The 2D wetting profiles induced by glycerol (a) and fluorinated-oil (b) droplets are plotted in the $z$-$r$ planes as the crosslinking density $k$ varies from $0.91~\%$ to $1.33~\%$. The dotted line indicates the liquid-air interfaces. The insets in both panels illustrate the balance of surface stresses at the contact points. The dashed lines in the panel (a) represent the theoretical predictions of the ridge profiles based on a continuum elastic model with the measured surface stresses determined by the Neumann's triangles.}
    \label{profiles}
\end{figure}

To measure the surface profiles of the substrates induced by wetting precisely, we deposited a layer of fluorescent nano-beads ($\sim 200$ nm) at the gel interfaces as the tracer particles. The beads had area densities less than $0.2\%$ so that, their influence on the surface properties of the gels was negligible \cite{Xu2017}. The fluorescent particles were imaged using a Leica-SP8 laser confocal fluorescent microscope with a 63x water immersion objective (N.A. $= 1.20$). For each droplet, we first obtained a stack image by scanning the focal plane vertically to cover the height of the wetting ridges. By locating the 3D positions of the nanoparticles, we can reconstruct the surface deformations with a spatial resolution around 200 nm.  Due to the axial symmetry of the wetting profiles with respect to the droplet center, we can further collapse the 3D profile azimuthally to a 2D plane. The resulting profiles induced by the glycerol and fluorinated oil droplets are presented as height ($z$) versus radial distance ($r$) in Figs.~\ref{profiles}a and b for various crosslinking densities. The plots are shifted to align onto the peaks of the wetting ridges.

\begin{figure}[t]
    \centering
    \includegraphics[width=0.48\textwidth]{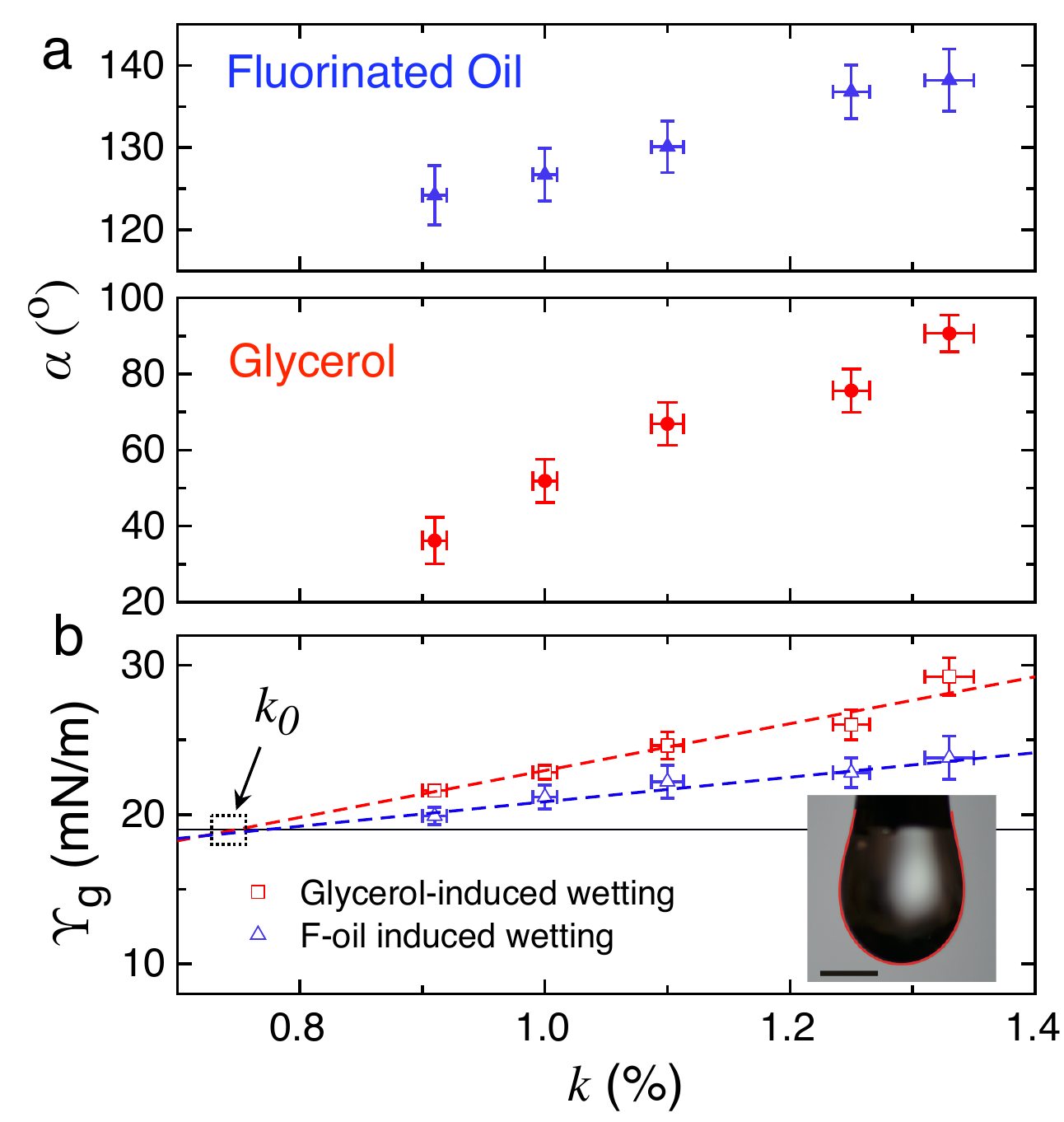}
    \caption{ {\bf Surface stress of soft gels with different crosslinking densities.} (a) The opening angles of the wetting profiles ($\alpha$) induced by both fluorinated oil (solid blue triangles) and glyceroal droplets (solid red diamonds) are plotted against the crosslinking $k$, respectively. The error bars in $\alpha$ indicate the standard deviations from the repeated measurements for at least five droplets in each experiment. (b). The surface stresses at the gel-air interfaces ($\Upsilon_g$) are measured for both glycerol (the hollow red squares) and fluorinated oil (the hollow blue triangles) droplets at various crosslinking densities ($k$). The dashed red and blue lines represent the linear extrapolations to the measured $\Upsilon_g(k)$ for the two deposited liquids. The black solid line indicates the surface stress of uncrosslinked polydimethylsiloxane (PDMS), $\Upsilon_0=19$ mN/m. Inset: a pendant droplet image of uncrosslinked PDMS where the scale bar is 1 mm long. The red solid line indicates the best fit to the droplet boundary.}
    \label{analysis}
\end{figure}

For the glycerol droplets wetting the soft gels, the overall geometry of the induced ridges varied greatly as $k$ decreased from $1.33\%$ to $0.91\%$. The height of the wetting ridges increased from $6~\mu$m to $55~\mu$m (Fig.~\ref{profiles}a), whereas the opening angle ($\alpha$) at the top of the ridges decreased correspondingly from $91$ to $36$ degree (Fig.~\ref{analysis}a). For a given crosslinking density of the substrate, we repeated the profile measurements with 5 to 10 glycerol droplets. The opening angle $\alpha$ was found to be independent of the droplet size. This generic profile near the top of ridges was similar to the Neumann's triangle in three-liquid contacts. The geometry is determined by the balance of surface stresses between different interfaces \cite{Style2013}.
Because the profiles were symmetric with respect to the glycerol-air interface (the dashed line in Fig.~\ref{profiles}e), the gel surface stresses on both sides were approximately equal, $\Upsilon_{gl} = \Upsilon_{ga} = \Upsilon_g$. Hence, we have the stress balance along the droplet surface, 
\begin{equation}
    2\Upsilon_g \cos({\alpha}/{2})= \Upsilon_{l}.
    \label{Neumann}
\end{equation}
Since the glycerol surface stress $\Upsilon_l = 40.7$ mN/m was independent of the crosslinking density $k$, the surface stress of the gels $\Upsilon_g$ can be determined exclusively by the opening angle $\alpha$ in Eq.~\ref{Neumann}. The hollow red squares in Fig ~\ref{analysis}b shows how $\Upsilon_g$ varied with the crosslinker ratio $k$ as the deposited liquid was glycerol. From $k=0.91\%$ to $k=1.33\%$, $\Upsilon_g$ increased by almost $50\%$, from $22$ mN/m to $30$ mN/m.  

The wetting profiles in Fig. \ref{profiles}a were compared to a continuum elastic model \cite{Style2012} by using the surface stresses ($\Upsilon_g$) and the shear moduli ($G_0$) of the soft gels with different crosslinking densities (see the Supplemental Materials for calculation details). As shown by the dashed lines in Fig. \ref{profiles}a, the theoretical predictions fitted well to the wetting profiles measured in experiments. Since $\Upsilon_g$ was determined by local geometries of the wetting ridges based on Eq. \ref{Neumann}, the nice agreement between the theory and experiments suggests that the surface stresses extracted from Neumann's triangle are consistent with the overall surface deformations induced by the wetting of glycerol droplets.

\begin{figure*}[t]
    \centering
    \includegraphics[width=170mm]{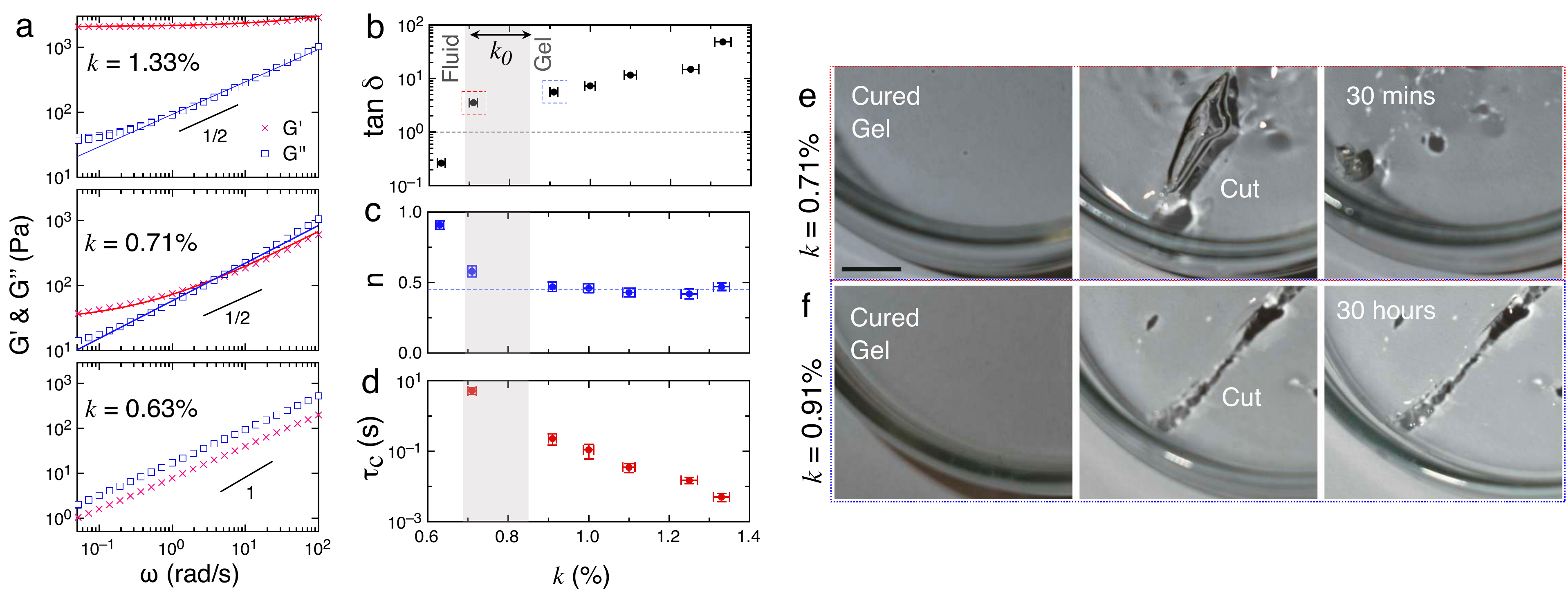}
    \caption{{\bf Material properties near the critical crosslinking density $k_0$.} (a) Plots of $G^\prime$ (red crosses) and $G^{\prime\prime}$ (blue squares) against the angular frequency $\omega$ for $k=1.33\%, 0.71\%$ and $0.63\%$, respectively. The solid lines indicate the best fits to the power-law rheological model, $G^\star(\omega) = G^\prime(\omega)+iG^{\prime\prime}(\omega) = G_0 (1+(i\omega\tau_c)^n)$. (b) A plot of $\tan{\delta} (= G^\prime/G^{\prime\prime})$  against the crosslinker ratio, $k$, at a frequency $\omega_0=0.1$ s$^{-1}$ where $\delta$ is the viscoelastic phase shift. The dashed line corresponds to the condition of $\delta=\pi/4$ or $G^\prime=G^{\prime\prime}$. The dashed red and blue boxes highlight the results for $k=0.71\%$ and $k=0.91\%$, respectively. (c) A plot of the fitted scaling index $n$ against the ratio of crosslinkers, $k$. The dashed blue line indicates $n=0.47$. (d) A plot of the fitted relaxation time scale $\tau_c$ against the ratio of crosslinkers, $k$. The grey regions in the panels (b)-(d) indicate the regime where $k_0 = (0.77\pm 0.08)$\%. (e) For the gels with a crosslinker ratio $k=0.71\%$, a cut at the interface is spontaneously healed within 30 minutes. (f) As $k$ increases to $0.91\%$, the cut at the interface remains.}.
    \label{Rheology}
\end{figure*}

A similar dependence on  the crosslinking density ($k$) was also observed in the wetting profiles of fluorinated oil droplets. As shown in the Fig. \ref{profiles}b, the surface deformations were small due to the low surface stresses of fluorinated oil. When increasing $k$ from $0.91\%$ to $1.33\%$, the opening angles ($\alpha$) increased moderately from 124 to 138 degrees (the solid blue triangles in Fig. \ref{Neumann}a). Due to the asymmetric wetting profiles in Fig. \ref{profiles}b, the surface stresses along the different arms of the ridges were not equal, $\Upsilon_{gl} \neq \Upsilon_{ga} =\Upsilon_g$. Therefore, we must consider the stress balances along both the $r$ and $z$ directions in this case to extract the surface stresses of the soft gels (as indicated by the inset of Fig.\ref{profiles}b). The hollow blue triangles in Fig ~\ref{analysis}b show how the surface stress at the gel-air interfaces, $\Upsilon_g$, varied with the cross-linker ratio $k$ when the deposited liquid was fluorinated oil. From $k=0.91\%$ to $k=1.33\%$, $\Upsilon_g$ increased gradually from $20$ mN/m to $24$ mN/m.

The change of gel surface stress $\Upsilon_g$ with crosslinking density for both deposited liquids is in contrast to the constant surface energy $\gamma_g$ found in contact angle measurements. In a control experiment, we measured the surface stress of uncrosslinked polydimethylsiloxane (PDMS) in air, $\Upsilon_0 = 19\pm 1$ mN/m, by using the pendant droplet method (Fig.~\ref{analysis}b inset) \cite{HANSEN1991}. Since the uncrosslinked PDMS is a Newtonian liquid, we assume its surface stress and surface energy to be equal, $\Upsilon_0 = \gamma_0$. Also because the surface energy of gels $\gamma_g$ does not vary with the density of crosslinkers, we further assume that $\gamma_g \approx \gamma_0 = \Upsilon_0 = 19$ mN/m. This value of $\gamma_g$ is indicated in Fig.~\ref{analysis}b by the black solid line. Meanwhile, as shown by both the red squares and blue triangles in the figure, we consistently found that $\Upsilon_g>\gamma_g$ for all of the samples we measured. The difference between $\Upsilon_g$ and $\gamma_g$ gradually diminished as the density of crosslinkers $k$ decreased. By applying linear extrapolations to the surface stress results measured with glycerol and fluorinated oil droplets in Fig.~\ref{analysis}b, we estimated that at a critical crosslinking density $k_0 = (0.77 \pm 0.08)\%$ the surface stress and surface energy of the soft gels became approximately equal, $\Upsilon_g  \approx \gamma_g$.

To understand the physical implications of the critical density $k_0$, we compared the surface properties with the bulk rheology of the soft gels near this transition. To characterize the viscoelasticity of the soft gels, we applied oscillation rheological tests to the materials at a small sweeping amplitude $\epsilon_0 = 1\%$. For each crosslinking density, we measured how the storage modulus $G^\prime$ and loss modulus $G^{\prime\prime}$ varied with the angular frequency $\omega$. Figure~\ref{Rheology}a shows typical viscoelastic spectra of the gels with $k=1.33\%$, $0.71\%$ and $0.63\%$, respectively. At $k=1.33\% > k_0$, the storage modulus was consistently greater than the loss modulus, $G^\prime(\omega) \gg G^{\prime\prime}(\omega)$. In this regime, solid-like gels formed properly. By contrast, as the ratio of crosslinkers decreased below $k_0$, the storage modulus became smaller than the loss modulus, $G^\prime(\omega) < G^{\prime\prime}(\omega)$, suggesting that  the viscous dissipation dominated the stress responses. At $k=0.63\%<k_0$, for example, the pre-mixed solution formed a gel by curing with difficulty and remained fluid at room temperature. Near the critical density ($k=0.71\%\approx k_0$), a very low shear modulus $G_0 = 28$  Pa can be measured. However, $G^\prime(\omega)$ remained consistently greater than $G^{\prime\prime}(\omega)$ only in the low frequency regime, $\omega < 1$ rad/s. For  $\omega > 1$ rad/s, the storage modulus and the loss modulus modulus became approximately equal, $G^\prime(\omega) \approx G^{\prime\prime}(\omega)$. As shown in Fig.~\ref{Rheology}b, this change in rheological behaviors is manifested in the plot of the phase delay ($\tan \delta$) against $k$ at a low frequency $\omega_0 = 10^{-1}$ rad/s. The phase shift $\delta$ is defined by the ratio of storage modulus and loss modulus, $\tan{\delta}=G^\prime/G^{\prime\prime}$. The grey area  in the plot indicates the transition regime near $k_0 = (0.71 \pm 0.08) \%$, and the black dashed line represents the critical condition of $G^\prime(\omega_0) =G^{\prime\prime} (\omega_0)$ at $k\approx k_0$. 

We further quantitatively compared the experimental results of $G^\prime$ and $G^{\prime\prime}$ with a power-law rheological model proposed by Chasset and Thirion \cite{Chasset_Thirion,Vega2001},
\begin{equation}
G^\star(\omega) = G^\prime(\omega)+iG^{\prime\prime}(\omega) = G_0 (1+(i\omega\tau_c)^n),
\label{power}
\end{equation}
where $n$ is a scaling index related to the network  and $\tau_c$ is an intrinsic relaxation time scale \cite{Long1996}. At $k > k_0$, the viscoelastic spectrum of the soft gels can be well described by Eq.~\ref{power}. As the ratio of crosslinkers decreases from 1.33\% to 0.91\%, the fitted scaling  index $n$ remained approximately constant around $n = 0.47$ (Fig.~\ref{Rheology}c) while the relaxation time scale increased substantially by two orders of magnitude (Fig.~\ref{Rheology}d), from $\tau_c = 4$ ms to $\tau_c = 0.23$ s. These results imply that the invariant scaling index $n\approx 1/2$ is a signature of a percolated network formed in the soft gels \cite{winter1986}. The associated viscoelastic relaxations, however, will slow significantly as the network softens. By contrast, due to the lack of a properly formed network, $G^\prime(\omega)$ and $G^{\prime\prime}(\omega)$ could no longer be fitted to Eq.~\ref{power} at a low crosslinking density, $k=0.63\%<k_0$. In this regime, although the relaxation time of the materials can not be determined, we observed an approximately linear scaling for the viscoelastic moduli against the angular frequency, $G^\prime\sim G^{\prime\prime} \sim \omega^{0.9}$.

Near the critical density $k_0 $, the materials showed unique mechanical properties.  At $k=0.71\%\approx k_0$, for example, the viscoelastic moduli  of the materials, $G^\prime(\omega)$ and $G{^{\prime\prime}}(\omega)$, can be well fitted to Eq. \ref{power} with a scaling index $n = 0.58$ and a long relaxation time $\tau_c = 5.3$ s. Since the Chasset-Thirion model has been widely used to explain the rheology of polymeric networks surrounded by free chains \cite{Vega2001, Batra2005,Karpitschka2015,MARTIN2008}, the results suggest that a percolated network with a small rigidity had already formed in this transition regime. However, the material surfaces remained to show liquid-like features in the response to interfacial fractures. As demonstrated in the images of Fig.~\ref{Rheology}e, a slight cut by a razor blade on the gel surfaces could be healed spontaneously in 30 minutes which was much longer than the viscoelastic relaxation time $\tau_c = 5.3$ s. This slow self-healing character indicates a high diffusivity of polymer chains at the interface \cite{Xu2021} while the bulk contained a weakly crosslinked network. For this reason, a nice coating of the fluorescent nano-particles at the interfaces became impossible when the crosslinking density was close to $k_0$. In a control experiment, as long as we increased the crosslinker ratio to $k=0.91\%$, a similar cut on gel surfaces in Fig.~\ref{Rheology}f  remained permanently. 

\section{Discussions and Conclusions}

Having observed the relationship between the surface properties and crosslinking densities of the soft gels, we now consider possible explanations to the results. Since the overall geometry of the wetting ridges can be well captured by a continuum mechanical model, the role of any possible phase separations \cite{Cai2021} of the excessive free chains near contact points should be insignificant in determining the surface stress. To further confirm this conclusion, we applied toluene treatments\cite{Xu2021} to partially remove the excessive free chains near the interfaces. For these partially dried samples, the increase of surface stress with the crosslinking density were still observed in the local measurements of the wetting profiles (see the supplemental materials).

Here we interpret the effects of the crosslinking density by considering the surface elasticity of the gels \cite{Xu2017}. As an example, the surface modulus $\Lambda_s$ for two different crosslinking densities, $k=1.33\%$ and $k=1.11\%$ respectively, can be estimated from the glycerol-induced wetting profiles. By locating the positions of fluorescent beads before and after depositing the glycerol droplets, we can determine their movements along the wetting ridges \cite{Boltyanskiy2017}.   By calculating the average surface deformations within the elastocapillary length, we can estimate the local surface strain ($\epsilon_s$) near the contact points \cite{Xu2017, Xu2018}. For $k=1.33\%$ and $k=1.11\%$, we approximately have $\epsilon_s\approx 0.11$ and $\epsilon_s \approx 0.17$. Considering a linear model for the surface stress $\Upsilon_g = \Upsilon_0 + \epsilon_s \Lambda_s$, we estimate that the surface modulus $\Lambda_s$ decreases from 93.1 mN/m to 37.5 mN/m as the crosslinker ratio $k$ changes from $1.31\%$ to $1.11\%$. Recent experiments involving spontaneous flattening on patterned gel surfaces also revealed a decrease of surface elastic modulus with gel stiffness \cite{Bain2021}. For $k\leq k_0=0.77\%$, we expect surface elasticity to vanish, $\Lambda_s=0$. In this regime, the average spacing between crosslinkers is too large to affect inter-polymer interactions. Thus, the material surface preserves the liquid-like feature, $\Upsilon_g = \gamma_g$. For $k>k_0$, surface elasticity appears ($\Lambda_s>0$) when the storage modulus becomes the dominating term for the  bulk rheology. As a result, the surface stress is increased by local wetting profiles and hence we  measured $\Upsilon_g>\gamma_g$ in experiments.

We further relate our results to the Shuttleworth effect, which was previously reported to explain the contact mechanics of soft silicone gels \cite{Xu2017, Jensen2018, Marchand2012,Grocum2018}. From the thermodynamics of any material interfaces, the relationship between surface stress  $\Upsilon_g$ and surface energy $\gamma_g$ can be expressed by the Shuttleworth equation,
\begin{equation}
\Upsilon_g(\epsilon_s)=\gamma_g(\epsilon_s)+\frac{\partial \gamma_g(\epsilon_s)}{\partial \epsilon_s}
\label{eq:shuttleworth}
\end{equation}
where $\Upsilon_g$ and $\gamma_g$ both depend on surface strains $\epsilon_s$ \cite{Cammarata1994}. Considering $\Upsilon_g= \Upsilon_0 + \Lambda_s\epsilon_s$ in Eq. \ref{eq:shuttleworth}, we can write the surface energy as $\gamma_g = \gamma_0 +  \Lambda_s\epsilon_s^2/2+O(\epsilon_s^3)$ where $\gamma_0 = \Upsilon_0$. The change of surface energy with surface strain hence follows a parabolic scaling, $\Delta \gamma = \gamma_g-\gamma_0 \sim \Lambda_s\epsilon_s^2/2$, while  
the change of surface stress scales linearly with the surface strain, $\Delta\Upsilon=\Upsilon_g-\Upsilon_0\sim\Lambda_s \epsilon_s$. The scaling difference between $\Delta\Upsilon$ and $\Delta\gamma$ can help to explain why we observed no change in the gel surface energy with crosslinking density $k$. Considering that $\epsilon_s$ is around $10\%$ due to the wetting ridges, the maximum increase in $\gamma_g$ is on the order of $0.5$ mN/m, smaller than the uncertainty of  the droplet surface stress measured in Fig. \ref{macroscopic}. Therefore, the contact angle measurements can not resolve the insignificant changes in gel surface energy during soft wetting.

Since surface stress ($\Upsilon_g $) and surface energy ($\gamma_g$) show strikingly different dependencies on crosslinking density, the macroscopic contact angle alone can not decide the surface properties of soft polymeric gels.  Our results show the importance of measuring the wetting profiles at different length scales to quantify the surface mechanics of soft gels fully. The distinction between the two regimes, $\Upsilon_g >\gamma_g$ and $\Upsilon_g = \gamma_g$, suggests a fundamental difference between the surfaces of soft gels and liquids. The crossover between the two regimes signifies a liquid-to-gel phase transition in the bulk. 

\section*{Conflict of interest}
The authors declare no competing financial interests. 

\section*{Acknowledgments}
We acknowledge Dr. Robert Style for useful discussions. We thank V. M. Vaisakh for helping us with the experiments with reduced free chains. We thank the support from the Early Career Scheme from the Hong Kong Research Grant Council (Grant No. HKUST 26309620).

\section*{Supplemental Materials}
\subsection{Measurements of the surface stresses of liquid droplets}
We measured the surface stress of Newtonian liquids (including glycerol, fluorinated oil and silicone oil) by numerically analyzing their droplet profiles. The images of both sessile and pendant droplets were taken by a digital Nikon D5600 camera equipped with a 105 mm macro-lens. The droplet edges were resolved by using the Canny boundary detector. The surface curvatures at the droplet interface ($\kappa$) result from the balance between the Laplace pressure and the hydrostatic pressure,               
\begin{equation}
2\Upsilon_l (\kappa - \kappa_0) = \pm \Delta\rho g z 
\label{eqn:stress}
\end{equation}
where $\kappa_0$ is the curvature at the apex and $\Delta \rho$ is the density difference between the droplet and the surrounding medium. Since we only measured droplets in air, the density difference $\Delta \rho$ can be replaced by liquid density $\rho$. The sign, $``\pm"$ , on the right-handed side of Eq. \ref{eqn:stress} is determined by whether the image was taken for a sessile droplet ($+$) or a pendant droplet ($-$). 

\begin{figure}[h]
\centering 
\includegraphics[width = 75 mm]{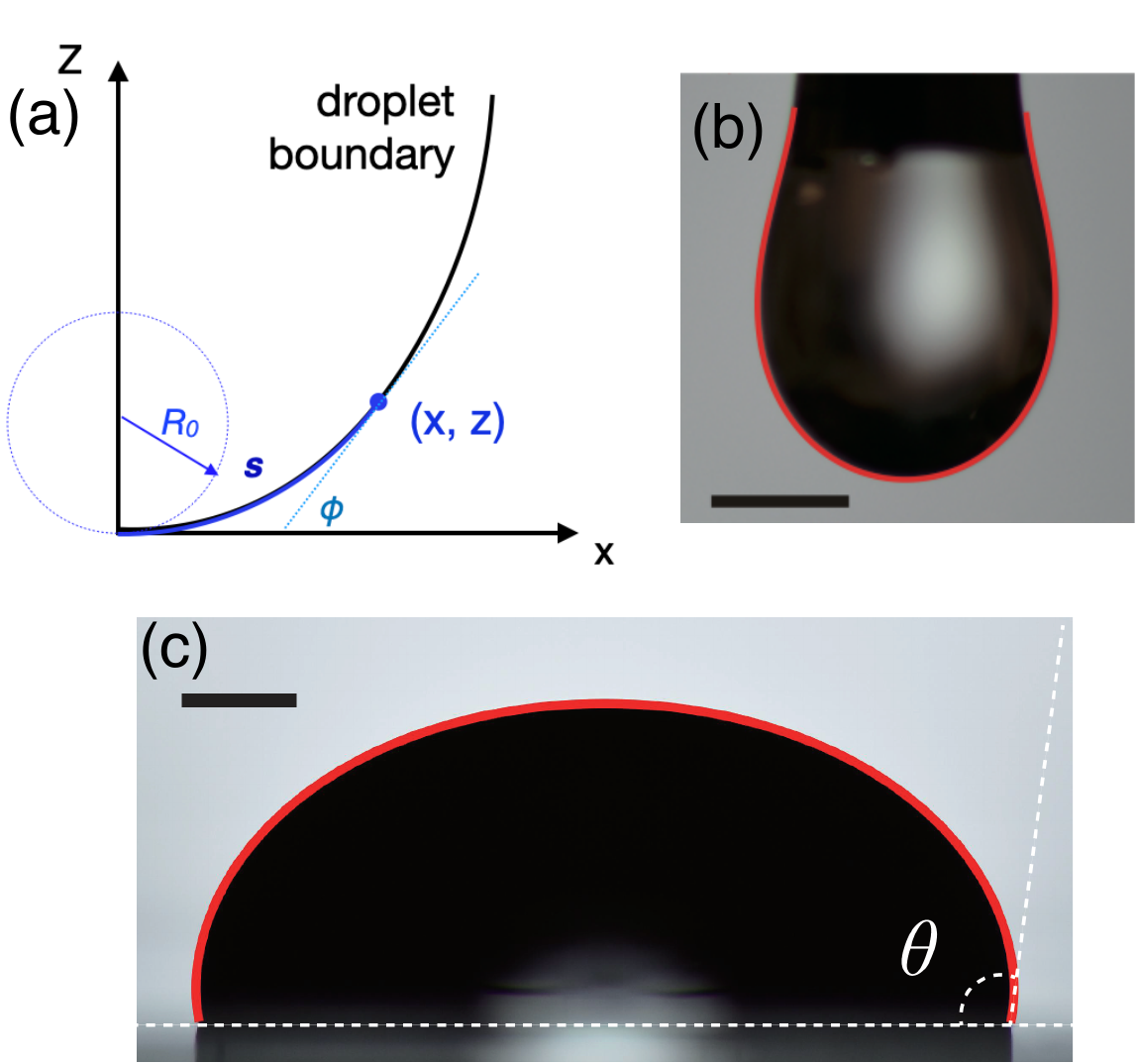}
\caption{{\bf a}. A schematic illustration of the droplet profiles defined by the parametric coordinates ($s,\phi $). {\bf b}. A pendent droplet with the fitted droplet boundary (the red solid line). Scale bar: 1mm. {\bf d}.  A sessile droplet with the fitted droplet boundary (the red solid line). Scale bar: 500 $\mu$m.}
\label{fig:droplet}
\end{figure}

Considering the axial symmetry of the droplet profiles, the boundary can be projected onto a 2D rz-coordinate plane (see Fig.\ref{fig:droplet} a). We use $s$ and $\phi$ to denote the arc length and the angle of the slope at the boundary interface, respectively. Therefore, $dx = ds \cos{\phi}$ and $dz =  ds\sin{\phi}$. Therefore, Eq. \ref{eqn:stress} can be rewritten in a parametric form,
\begin{equation}
\begin{aligned}
\left\{
        \begin{array}{lr}
          \mathrm{d}X = cos \phi~\mathrm{d}S\: \: \: \:  &  \\

          \mathrm{d}Z = sin \phi~\mathrm{d}S \: \: \: \: \: \: \: \: \: \: \: \: \: \: \:  &  \\
          
          \mathrm{d}\phi = (2-{sin\phi}/{X} \mp \beta_0Z)\mathrm{d}S, &  \\ \beta_0 = {\Delta\rho g R_0^2}/{\Upsilon_l}
        \end{array}
\right.
\end{aligned}
\label{eqn:parametric}
\end{equation}
where the coordinates $(x,z,s)$ are normalized to the dimensionless variables $X=x/R_0, Y=y/R_0$ and $S = s/R_0$. We can calculate the droplet profile  numerically through the iterations of Eq. \ref{eqn:parametric}. The droplet surface stress $\Upsilon_l$ is considered to be successfully determined when the numerical result fits well with the imaged droplet profile. Equations \ref{eqn:stress} and \ref{eqn:parametric} can be applied to both pendant (Fig. \ref{fig:droplet}b) and sessile droplets (Fig. \ref{fig:droplet}c). 

\begin{figure}[h]
\centering 
\includegraphics[width = 85 mm]{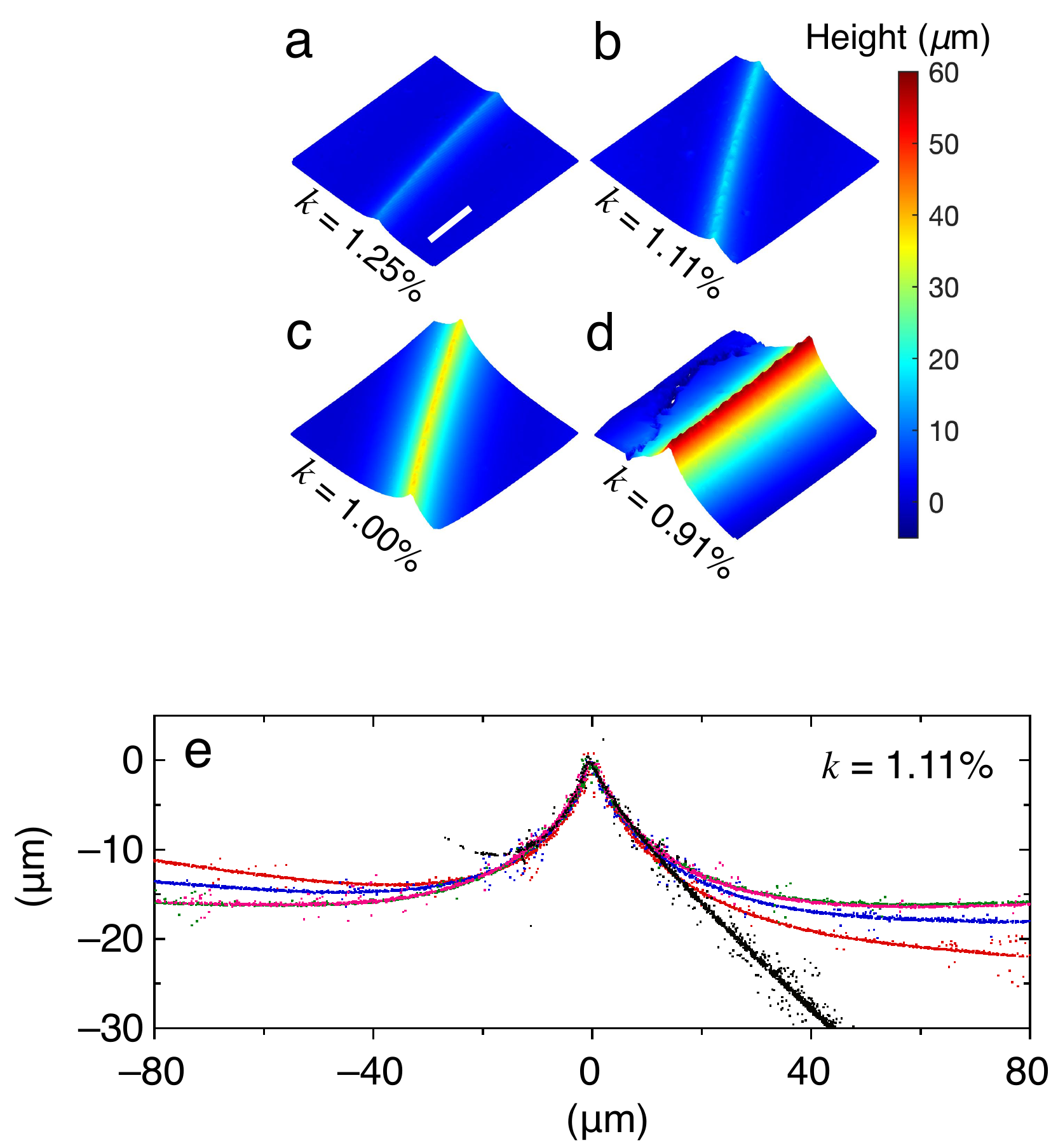}
\caption{{\bf a - d}. Reconstructed three-dimensional surface profiles from confocal microscope measurements of the gel substrates for $k=1.25·\%, 1.11~\%, 1.0~\%,$ and $0.91`\%$, respectively. The color bar indicates deformations along $z$. The scale bar in the x-y plane indicates 50 $\mu$m. {\bf e}. The collapsed wetting profiles induced by different sizes of droplets for $k = 1.11\%$.}
\label{fig:interference}
\end{figure}

\subsection{Confocal microscopy imaging of the local wetting profiles} 
To place the tracers on the surfaces of soft silicone gels, we deposited a solution droplet containing 200-nm beads on the surfaces of the cured gels for at least 3 h. During the coating process, the entire sample was sealed properly to avoid any evaporation. A fraction of the beads in the solution diffused to the gel and adhered to the interface. As a result, a layer of nicely coated fluorescent beads was left on the gel surfaces when the solution was removed. In the soft wetting experiments, we used a spinning-disk laser confocal microscope (Lecia SP8) to image the region close to the contact point. By locating the fluorescent beads in 3D, we can reconstruct the wetting profiles by using our previously developed MATLAB codes (as shown in Fig. \ref{fig:interference} a-d). Due to axial symmetry of the droplet geometry, all of the wetting profiles are collapsed to the 2D $r-z$ plane. For each gel substrate, we varied the droplet radius and collapsed the wetting profiles near the contact point to confirm the validity of the Neumann's triangle (Fig. \ref{fig:interference} e). 

\begin{figure}[t]
\centering 
\includegraphics[width = 85 mm]{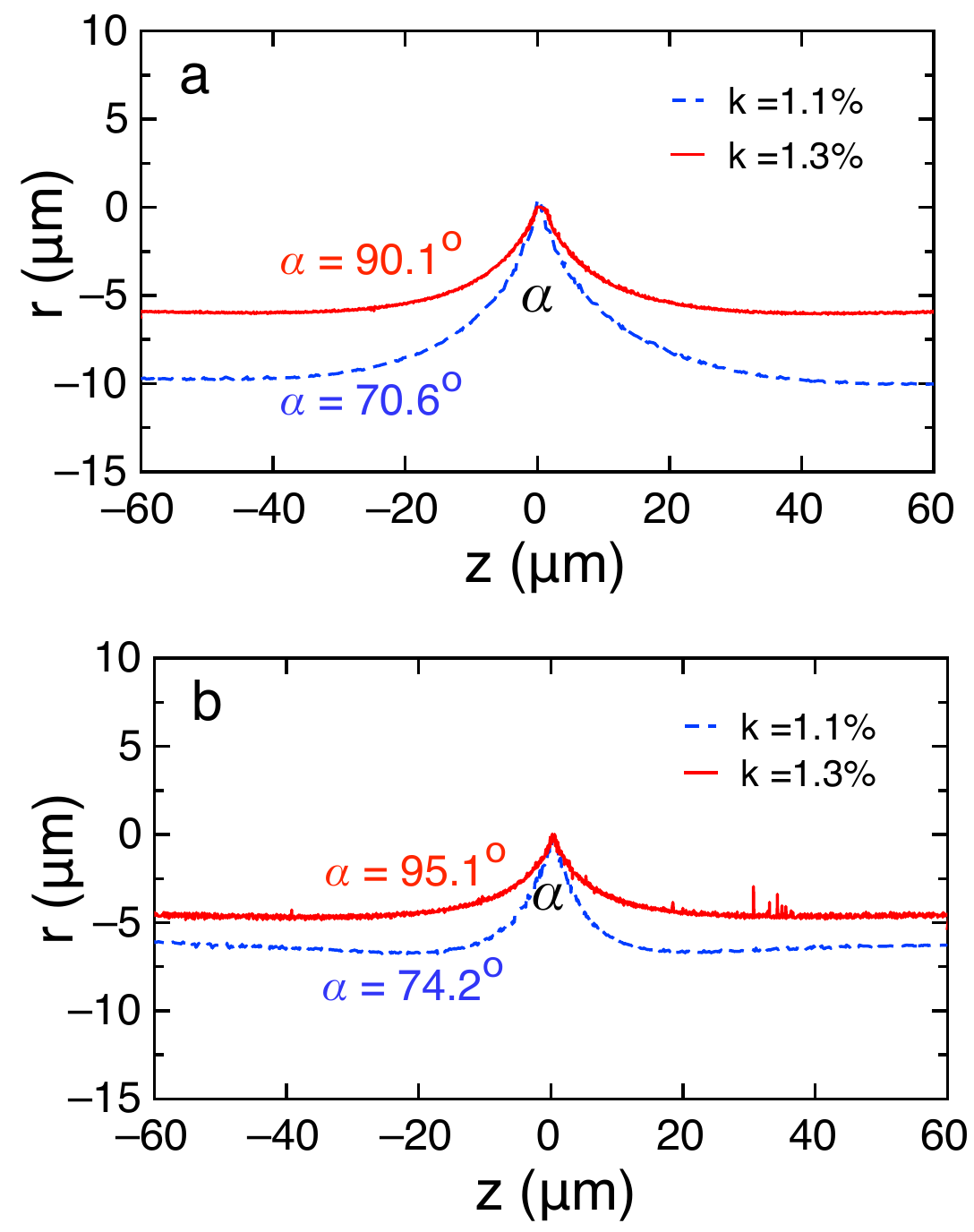}
\caption{The wetting profiles before (a) and after (b) the swelling and drying procedures for two different crosslinking densities $k=1.1\%$ (blue dashed line) and $k=1.3\%$ (red solid line), respectively. In both panels (a) and (b), we observed the same trend that the opening angle $\alpha$ increases with the crosslinking density $k$.}
\label{fig:swell}
\end{figure}

\subsection{Wetting profiles on the gels with reduced free chains}
To investigate the role of free chains in surface stress measurements, we performed control experiments on gel surfaces with reduced amounts of uncrosslinked polymers. The samples were prepared through the following three steps. First, a cured gel substrate coated on a glass slide was submerged in a $50\%$ toluene solution mixed with $50\%$ ethanol for 24 h. The process greatly swelled the gel networks and extracted uncrosslinked polymers by osmotic pressure near the gel interfaces. Second, we removed the surrounding solution containing the free chains and then waited for another 24 h for the toluene and ethanol to dry out. Third, we repeated the first two steps again to further reduce the amount of free chains near the interfaces. This treatment  decreased the substrate mass by approximately $15\%$ without increasing the surface roughness. Figures \ref{fig:swell} a and b show the wetting profiles before and after the treatment for $k=1.3\%$ and $k=1.1\%$, respectively.

\subsection{A Continuum elastic model for soft wetting}
In this work, we extended the linear elastic theory proposed by Style, et al. (Ref.[29] in the paper) to soft wetting with a given contact angle $\theta$. The calculation assumes that $\Upsilon_{gl}\approx \Upsilon_{ga} = \Upsilon_g$, which is consistent with the experiments on glycerol droplets. The governing equations of the displacement and stress fields, $(u(r,z))$ and $(\sigma(r,z))$, of the substrate are
\begin{align}
& (1-2\nu) \nabla^2 u + \nabla(\nabla\cdot u)=0,\\
& \sigma = \frac{e}{1+\nu} [\frac{1}{2} ((\nabla u)^T + \nabla u)+\frac{\nu}{1-2\nu}(\nabla\cdot u) {\bf I}].
\end{align}
Considering the boundary conditions due to the gel surface stress and liquid surface tension
\begin{align}
& \sigma_\Upsilon = \Upsilon_g \frac{1}{r} \frac{\partial}{\partial r}  (r\frac{\partial u_z}{\partial r})\hat{z}, \\
&\notag t(r,z=h)= \gamma_l \sin{\theta} \delta (r-R\sin\theta) \hat{z} - \frac{2\gamma_l}{R} H(R\sin\theta - r) \hat{z} \\
&  - \gamma_l\cos\theta\delta(r-R\sin\theta) \hat{r}, 
\end{align}
we can solve equations by applying Hankel transformations to both the displacement $u_z(r,z)$ and stress fields $\sigma(r,z)$. As a result, the surface profile $u_z(r,z=h)$ can be written as
\begin{align}
&\notag u_z(r,h) = 
\int_0^{+\infty} ds\   {\gamma_l} s J_0(sr) \times \\ 
& \notag ({J_1(sR\sin\theta)} s (\nu+1) \cos\theta  \\ 
& \notag\times \left(2 h^2 s^2+(2 (5-4 \nu) \nu-3) \cosh (2 h s)+2 \nu (4 \nu-5)+3\right)\\ 
   &\notag+2  {J_0(sR\sin\theta)} R s \left(\nu^2-1\right) \sin\theta ((4 \nu-3) \sinh
   (2 h s)+2 h s)\\ 
   &\notag-4 {J_1(sR\sin\theta)} \left(\nu^2-1\right) ((4 \nu-3)
   \sinh (2 h s)+2 h s)) \\
    &\notag/(s^2 (E \left(2 h^2 s^2+4 \nu
   (2 \nu-3)+5\right)+E (3-4 \nu) \cosh (2 h s)\\ 
   &+4 {\Upsilon_g} h s^2
   \left(\nu^2-1\right)+2 {\Upsilon_g} s (\nu-1) (\nu+1) (4 \nu-3) \sinh
   (2 h s))).
   \label{eq:integral}
\end{align}
The dashed lines in the Fig. 2a of the main manuscript were calculated by using Eq. \ref{eq:integral} with the experimental parameters for various cross-linking densities. The Poisson ratio was chosen as $\nu = 0.46$ in the calculations, which is consistent with the results obtained from our previous measurements (see Ref. [22] in the main manuscript).

\bibliography{reference_qx.bib}
\bibliographystyle{rsc}

\end{document}